\documentclass{PoS}
\usepackage{subfigure}
\def\la{\mathrel{\mathpalette\fun <}}
\def\ga{\mathrel{\mathpalette\fun >}}
\def\fun#1#2{\lower3.6pt\vbox{\baselineskip0pt\lineskip.9pt\ialign{$\mathsurround=0pt#1\hfil##\hfil$\crcr#2\crcr\sim\crcr}}}

\title{Indirect searches of dark matter }

\ShortTitle{Indirect searches of dark matter }

\author{\speaker{Le Zhang}\\%
       University of Hamburg, II. Institute for Theoretical Physics\\
       E-mail: \email{lezhang@mail.desy.de}}

\abstract{If dark matter decays or annihilates into electrons and positrons, it can affect radiation and cosmic-ray backgrounds. We review a novel, more general analysis of constraints on decaying dark matter models, by introducing the response functions based on the current radio, gamma-ray and positron observations. Constraints can be simply obtained by requiring the convolution of the response functions with actual decay spectrum of electrons and positrons smaller than the product of decay lifetime in $10^{26}$s and mass in $100$GeV. The response functions just depend on the astrophysical inputs such as the propagation model, but not on the microscopic decay scenario. 

Moreover, an anisotropy analysis of the full-sky radio emissions to identify the extragalactic dark matter annihilation is shown. We discuss the angular power spectra of the cosmological synchrotron emission from dark matter annihilations into electron positron pairs and compare them with astrophysical backgrounds and Galactic foregrounds. We find that the angular power spectrum of radio fluxes at around GHz frequencies and in the range of $200\la l\la3000$ opens a optimal window to disentangle the dark matter signals from common astrophysical backgrounds.}

\FullConference{Identification of Dark Matter 2010\\
		 July 26 - 30 2010\\
		 University of Montpellier 2, Montpellier, France}

\begin{document}

\section{Introduction}
The most popular class of dark matter models is the Weakly Interactive Massive Particles (WIMPs). Within standard cosmology, due to effective self-annihilation in the early Universe, dark matter can reproduce the correct relic density if the WIMPs are thermally ``freeze-out''. In addition, if a conserved quantum number is slightly violated, dark matter will become unstable and decay happens. The current astrophysical and cosmological observations require the decay of dark matter with lifetimes around and above $\tau_\chi\simeq \mathcal{O}(10^{26})$s. Within this framework, dark matter decays or annihilations can give rise to observable fluxes of gamma-rays, electrons, positrons, neutrinos, and even some antimatter such as anti-protons and positrons. Motivated by the recent results from cosmic-ray experiments~\cite{Abdo:2009zk,Adriani:2008zr,Adriani:2008zq,chang:2008zzr}, dark matter annihilating or decaying into leptonic final states offers an possible explanation to the PAMELA positron excess and is also compatible with the Fermi-LAT electron-positron data. The yielding electrons and positrons emit synchrotron radiation in the magnetic fields of galaxies which can be detected in the radio band, and they also produce gamma-rays through inverse Compton scattering (ICS) off the low energy background photons in an interstellar radiation field (ISRF) and through bremsstrahlung emissions due to the interaction with an ionized interstellar medium (ISM). These secondary radiations could provide indirect evidence for the particle nature of dark matter. 

In this paper, I review recent theoretical work related to spectral and anisotropy analysis of the dark matter decay or annihilation. The content presented is based primarily on the papers~\cite{Zhang:2008rs,Zhang:2009pr,Zhang:2009ut}. In Sect.~2, I show a universal method to derive constraints on dark matter decaying models. I introduce response functions based on three observables, namely the positron flux on Earth, the synchrotron radiation and the gamma-ray flux. Constraints can then be simply obtained by requiring the convolution of the response functions with a given dark matter decay spectrum smaller than a quantity related to the mass and lifetime of dark mater. In Sect.~3, I point out the anisotropic radio emission from the extragalactic dark matter annihilating into electrons and positrons can be regarded as a unique signature for disentangling dark matter siganls for astrophysical contributions.  I summarize my conclusions in Sect.~4.
\section{Spectral signatures of Galactic dark matter decay}
Usually constraints derived in the literature always focus on particular dark matter models with given decay spectra and branching ratios into the final state products. In contrast to these model-dependent results, in Ref.~\cite{Zhang:2009pr,Zhang:2009ut}, we propose a general method to calculate the signals resulting from electrons and positrons produced in dark matter decays and obtain the relevant constraints from astrophysical observations. Based on the fact that the propagation equation is linear with respect to the electron density, the injected electron energy thus evolves independently. With a finite number of numerical simulations for injected positrons or electrons at different energies, we can construct a numerical {\it response function} of the ratio of predicted dark matter signals to astrophysical backgrounds. The strongest constraint requires a special direction or patch in the sky for which the ratio of predicted and observed signal is maximal. The response function only depends on astrophysical parameters such as the cosmic-ray propagation model and the dark matter halo profile, but {\it not} on the specific decay model.

\begin{figure}[hbt]
\centering
 
\includegraphics[height=5 cm]{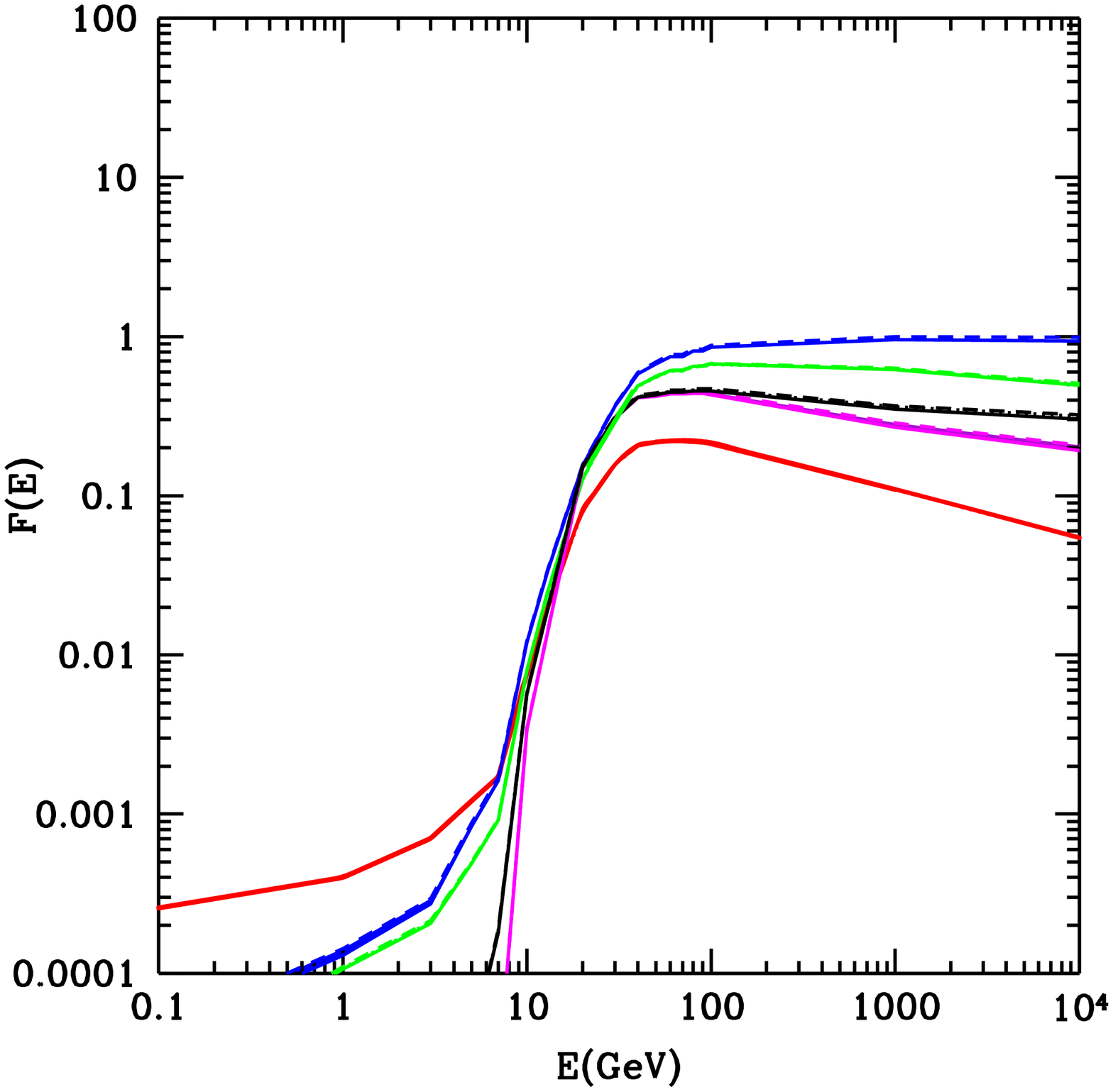}
\includegraphics[height=5 cm]{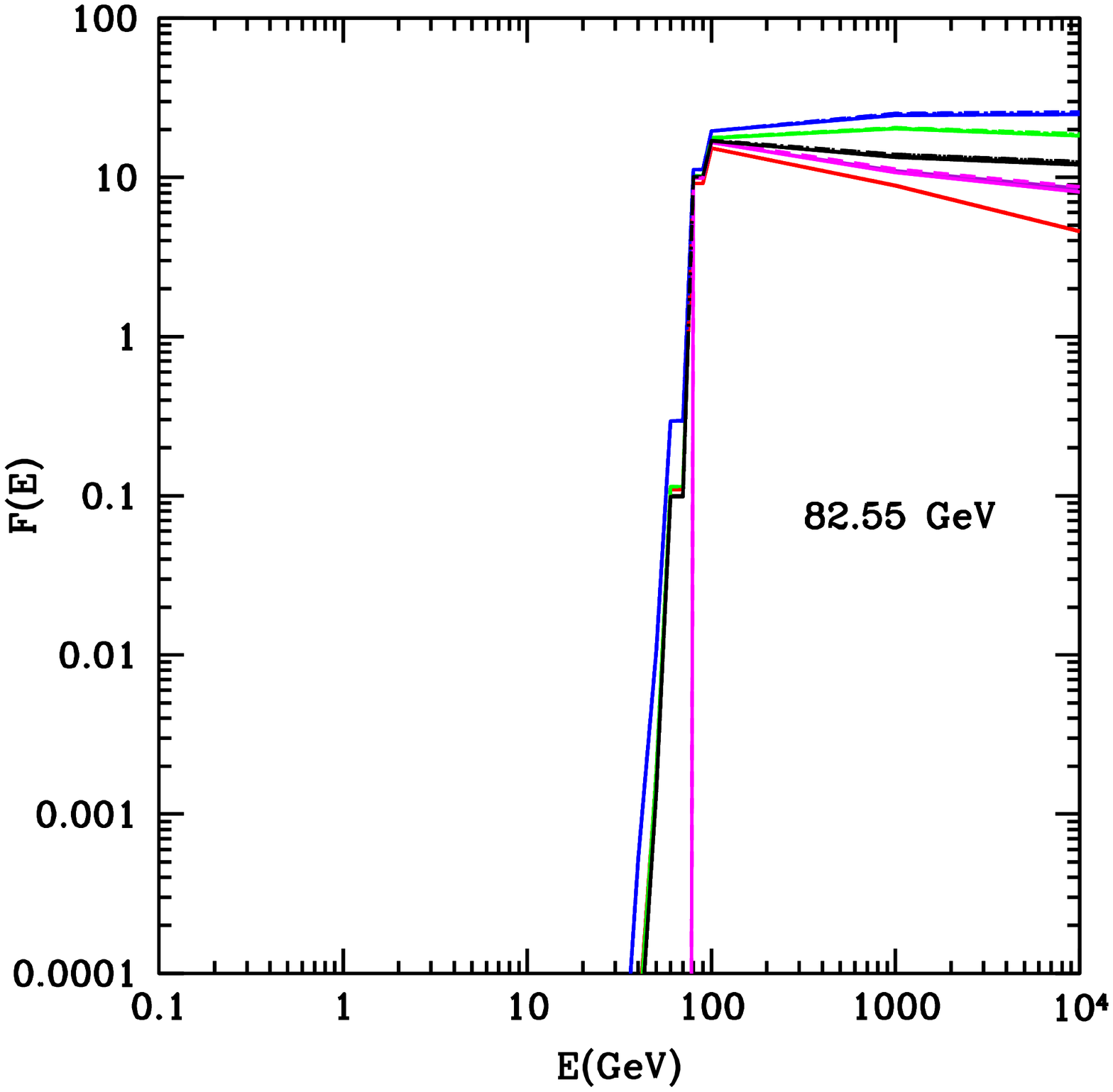}

\caption{The model dependence of the response function is shown. 
The response functions based on the observed radio flux (left) at 23 GHz and the positron flux (right) for the highest PAMELA energy bin are given. The color curves denote the response functions for different propagation models. Figures are from Ref.~\cite{Zhang:2009pr}
}
\label{fig:response}
\end{figure}

\begin{figure}[htp]
 \centering
 \includegraphics[height=5 cm]{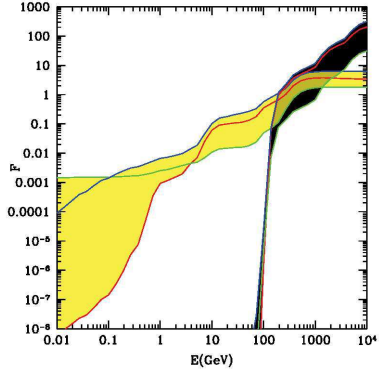}
 \caption{The propagation model dependence of the  response function
 based on the gamma-ray observation by Fermi-LAT~\cite{Dobler:2009xz} at energy range
 $0.5-1$ GeV (yellow band) and
 $100-300$ GeV (black band).  The
 width of the bands represents the variation within the propagation models. Figure is from Ref.~\cite{Zhang:2009ut} }
 \label{fig:error}
\end{figure}

Figures~\ref{fig:response}~and~\ref{fig:error} present the response functions for the radio, positron and gamma-ray fluxes, while showing the dependence on the propagation models. Ref.~\cite{Zhang:2009pr,Zhang:2009ut} point out that the largest uncertainties of the response functions at injection energys above 100GeV come from the poor knowledge about the height of the diffusion zone. The corresponding uncertainties can reach one order of magnitude. The other cosmic-ray propagation parameters, in particular from the strength of the reacceleration introduce the uncertainties at the energies below around 10 GeV.

Robust constraints can be obtained in terms of convolving the response functions $F(E;E_0)$ for the observed fluxes at energy $E$ with the specific decay spectrum into electrons and positrons $dN_e/dE_0$, requiring
\begin{equation}
\int_{m_e}^{m_X} d E_0\, F(E;E_0)\frac{dN_e}{dE_0}\leq
\left(\frac{\tau_X}{10^{26}\,{\rm s}}\right)\left(\frac{m_X}{100\,{\rm GeV}}\right)\,.
\end{equation}

Applying our method to provide model independent constraints on concrete decay modes, we found that the strongest constraints on injection spectrum at low energies ($\la1000$GeV) come from the positron data due to its lowest background and at high energies ($\ga1000$GeV) from gamma-ray data. The radio data always provide a relatively weaker constraining power.

\section{Radio anisotropy signatures of extragalactic dark matter  annihilation}
The angular power spectrum provides a statistical measure of fluctuations as a function of angular scale, which can be used to identify dark matter signatures from astrophysical backgrounds. Given that the extragalactic radio background is much lower than other radiation backgrounds, the radio observations would have a better sensitivity in search for the signatures of the dark matter annihilation. In Ref.~\cite{Zhang:2008rs}, we calculated the intensity and angular power spectrum of synchrotron emissions emitted by electrons and positrons produced by dark matter annihilation in the magnetic field of cosmological dark matter halos. The magnetic field in dark matter halos is adopted to be $B=10\,\mu$G which is realistic value since most annihilations occur in the central regions where magnetic fields are somewhat larger than typical average galactic fields of a few micro-Gauss. To evaluate the dark matter signals, we adopt  $m_\chi =100$ GeV and  $\left\langle\sigma v\right\rangle=3\times10^{-26}\,{\rm cm}^3$/s. we assume that the average
total number of electrons and positrons per annihilation is $Y_e\simeq10$, and that a boost factor $A_b\simeq10$ from substructures confirmed by recent numerical simulations.

\begin{figure}[htp]
\centering 
\includegraphics[height=7 cm]{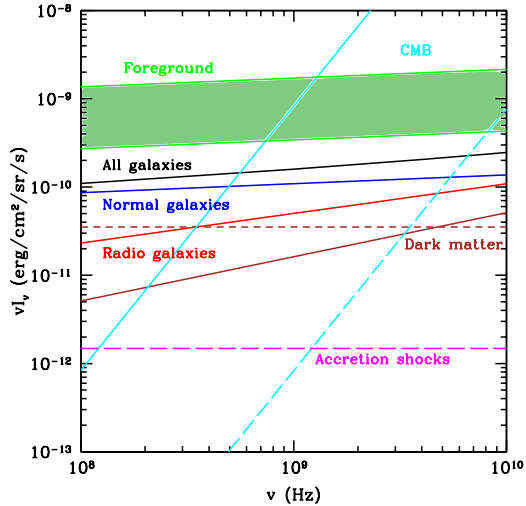}
\caption{Comparison of the intensity of diffuse radio emission from various components, including normal galaxies (blue curve), radio galaxies (red curve), radio and normal galaxies combined (black curve),  galaxy cluster shocks (magenta curve), and dark matter annihilations (brown curves). Also shown is the
CMB background (cyan solid curve) as well as its subtractable part (dotted cyan curve), and the Galactic foreground (green band). Details are given in Ref.~\cite{Zhang:2008rs}.}
\label{fig:radio_bg}
\end{figure}

Following the above assumptions,  in Fig.~\ref{fig:radio_bg}, the different contributions to the average diffuse radio backgrounds are shown. We can find there is an optimal window at frequencies $\nu\simeq\,2$GHz for detecting dark matter annihilation signals. Above these frequencies the contribution from CMB increases rapidly and below these frequencies the dark matter signal decreases whereas the astrophysical backgrounds tend to be flat. In any case, the Galactic foregrounds always contaminate the radio sky over the whole frequency range interested here. Given that the Galactic foregrounds have a smooth directional dependence, we, therefore, consider in the following the angular power spectra of the radio sky  to extract dark matter signatures.

\begin{figure}[htp]
\centering
\includegraphics[height=7 cm]{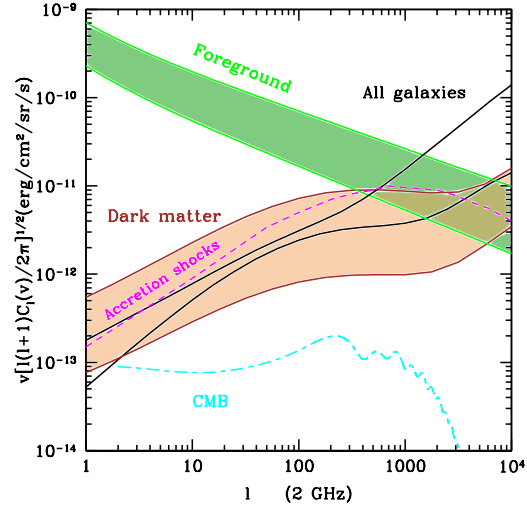}
\caption{Angular power spectra of the radio sky at 2 GHz compared with various estimates of the
Galactic foreground (green shaded region), the CMB (cyan curve), the all galaxies (black), the intergalactic shocks (magenta curve) and the dark matter annihilations (brown band). Details are given in Ref.~\cite{Zhang:2008rs}.}
\label{fig:cl_sum}
\end{figure}
 
Since the angular fluctuations are proportional to the mean intensity, we thus adopt the 2 GHz as our optimal frequency to calculate the angular power spectrum.  The cosmological background power spectra $C_l$ from various origins at 2 GHz are shown in Fig.~\ref{fig:cl_sum}. We can find that for dark matter annihilation the distribution of $l(l+1)C_l$ is nearly flat for multiples $200\la l\la2000$. At smaller $l$ the power spectrum is dominated by Galactic foregrounds and at larger $l$ dominated by galaxies due to the Poisson noise. If we now scale the dark matter signal with a additional boost factor of 10, the rather flat power spectrum of the dark matter annihilation signal would dominate over other cosmological backgrounds for $200\la l\la3000$, and should become distinguishable from the Galactic foreground. We can assert that the future radio observations are sensitive to 
\begin{equation}\label{eq:F_dm}
  \left(\frac{A_b}{10}\right) \left(\frac{Y_e}{10}\right)
  \left(\frac{\left\langle\sigma v\right\rangle}{3\times10^{-26}{\rm cm}^3{\rm s}^{-1}}\right)
  \left(\frac{100 {\rm GeV}}{m_X}\right)^2\left(\frac{10\,\mu{\rm G}}{B}\right)^{1/2}\ga 10\,.
\end{equation}

\section{Summary}
Dark matter decaying or annihilating into electrons and positrons can affect astrophysical backgrounds. In light of the recent experimental observations of the radio, gamma-ray and positron fluxes, we introduce useful response functions that can be applied to constrain any decaying dark matter models. Robust constraints can be obtained by the convolution of the response functions with a given DM decay spectrum and requiring that the result should be smaller than the product of DM mass and lifetime in suitable units. The different observations provide complementary bounds on the properties of dark matter.

In addition, anisotropies in diffuse radio emission can be a powerful diagnostic of the properties of contributing sources. For the radio backgrounds around $\simeq2$GHz  and its angular power spectrum for multipoles $200\la l \la 3000$, the annihilation signals can be disentangled from astrophysical backgrounds. The future radio observations such as SKA can provide valuable information on the nature of dark matter.

\section*{Acknowledgments}
I would like to thank the organizers of {\it Identification of Dark Matter 2010} for an perfect conference. I am also very
grateful to Luca Maccione, Javier Redondo,  G\"unter Sigl,  Christoph Weniger, for the rich collaboration and discussion.

\end{document}